\documentclass{article}

\usepackage[english]{babel}

\usepackage[letterpaper,top=2cm,bottom=2cm,left=3cm,right=3cm,marginparwidth=1.75cm]{geometry}

\usepackage{amsmath}
\usepackage{natbib}
\usepackage{bm}
\usepackage{graphicx}
\usepackage[colorlinks=true, allcolors=blue]{hyperref}
\usepackage{natbib}
\usepackage{multirow}
\usepackage{tabularx}
\usepackage{rotating}
\usepackage{caption}
\usepackage{subcaption}
\usepackage{nicematrix}
\usepackage{enumitem}
\usepackage{xcolor}

\newcommand{\changed}[1]{{#1}}

\title{Creating Treatment and Component Hierarchies in Component Network Meta-Analysis}
\author{Augustine Wigle, Audrey B\'{e}liveau, Adriani Nikolakopoulou, Lifeng Lin}
\date{}
\begin{document}
\maketitle
\newcommand{\sset}{\mathcal{S}}
\newcommand{\ssetf}{\mathcal{S}^{*}}
\newcommand{\tset}{\mathcal{T}}

\newcommand{\bbeta}{\bm{\beta}}
\newcommand{\X}{X}
\newcommand{\y}{\bm{y}}
\newcommand{\V}[1]{\bm{V}_{#1}}
\newcommand{\hbbeta}{\hat\bbeta}
\newcommand{\hbeta}[1]{\hat{\beta}_{#1}}
\newcommand{\hdelta}[1]{\hat{\theta}_{#1}}
\newcommand{\hdeltam}[1]{{\hdelta{#1}}^{(m)}}

\newcommand{\E}{\text{E}}
\newcommand{\rank}{\text{rank}}
\newcommand{\rankm}{\text{rank}^{(m)}}
\newcommand{\zero}{\bm{0}}
\newcommand{\hVar}{\widehat{\text{Cov}}}
\newcommand{\sucra}{\text{SUCRA}}
\newcommand{\pik}{p_{ik}}
\newcommand{\Pbar}{\overline{P}}
\newcommand{\Pbest}{P_{best}}

\newcommand{\BB}{\mathcal{B}}
\newcommand{\CC}{\mathcal{C}}
\newcommand{\MM}{\mathcal{M}}

\newcommand{\uu}{\bm{u}}
\newcommand{\UU}{\bm{U}}

\newcommand{\summ}{\sum_{m=1}^M}
\newcommand{\sumneq}{\sum_{\substack{j \in \ssetf \\j \neq i}}}

\newcommand{\subsubsubsection}[1]{\noindent{\textbf{#1}}}

\section{Introduction}

Network Meta-Analysis (NMA) is a method of synthesizing data from randomized controlled trials to simultaneously compare treatments or interventions that have not necessarily been compared head-to-head. Some networks involve multi-component treatments, for example, an intervention for treating depression could involve one or more antidepressant drugs and psychotherapy modalities. A review found that the incidence of multi-component treatments in networks is increasing, necessitating methods tailored to this type of intervention \citep{petropoulou_review_2021}. 
Component Network Meta-Analysis (CNMA) \citep{welton_mixed_2009, rucker_network_2019} is an extension of NMA that enables estimation of multi-component treatment effects and individual component effects, such as the effect of antidepressant drug A versus placebo, and antidepressant drug A + cognitive behavioural therapy versus placebo. In addition to estimating component effects, CNMA can be used to synthesize studies which do not share common treatments, but share one or more common components. 

Both Bayesian and frequentist approaches to CNMA have been proposed \citep{welton_mixed_2009, rucker_network_2019, efthimiou_bayesian_2022, wigle_bayesian_2022}. In a Bayesian model, estimation and inference is performed via Markov Chain Monte Carlo (MCMC) sampling, which provides many samples from the posterior distribution of the parameters. The frequentist approach to CNMA typically uses a weighted least squares approach, which provides point estimates for parameters along with an estimated variance-covariance matrix. 

A key output of an evidence synthesis and NMA is a hierarchy of the competing treatments \citep{papakonstantinou_answering_2022}. Treatment hierarchies provide a simple summary of the results of an analysis, although they involve a loss of information compared to a league table or forest plot, so should not be interpreted in isolation \citep{hutton_prisma_2016, salanti_introducing_2022}. Ranking metrics that summarise the estimated relative effects of each treatment compared to its competitors and their corresponding precisions are calculated and ordered to create treatment hierarchies. Different ranking metrics correspond to different criteria for determining preference among the competing treatments \citep{salanti_introducing_2022}. In standard NMA, hierarchies can answer questions like, ``what is the hierarchy of the competing treatments in terms of the probability of giving the best outcome value?".

\changed{In standard NMA of connected networks, the set of relative effects that can be estimated is simply those between the set of observed treatments, and a framework for creating treatment hierarchies through the calculation of ranking metrics for all observed treatments has been established \citep{salanti_introducing_2022}. However, in CNMA, the structure of the design matrix changes what is estimable, and may enable estimation of other relative effects beyond the observed treatments. Additionally, CNMA can enable synthesis of sparse, disconnected networks, but the relative effects of observed treatments are not necessarily meaningful in such networks. A graphical approach to understanding which component effects can be estimated from a given CNMA model was introduced recently \citep{li_graphical_2023}, but the question of what exactly can be uniquely estimated from a CNMA model has not yet been definitively answered in the literature. 

The set of parameters that can be meaningfully estimated in a CNMA model, and thus suitable for use in creating a hierarchy, is related to the identifiability of the model. A model is identifiable if different parameter values correspond to different models \citep{cole_determining_2010, cole_parameter_2021}. In the context of CNMA, a model is identifiable when different sets of parameter values correspond to different mean relative effects.
This definition applies to the likelihood, and thus can be applied in both Bayesian and frequentist models, although alternative definitions that also depend on the prior distribution are available in a Bayesian context \citep{kadane_role_1975,cole_parameter_2021}. If a model is non-identified from a classical perspective, there is a flat ridge in the likelihood function, and if the likelihood is paired with uninformative priors in a Bayesian model, the posterior will also have a flat ridge \citep[pg. ~128]{cole_parameter_2021}. Since the use of diffuse priors is standard in Bayesian CNMA, the likelihood-based definition of identifiability is suitable in this context for both Bayesian and frequentist models.}

Although CNMA is an increasingly relevant evidence synthesis technique, methods and guidelines for the creation of treatment and component hierarchies from CNMA have not yet been established.
CNMA may be able to answer additional hierarchy questions beyond what is possible in standard NMA, such as ``which component has the best performance out of the observed components?". Graphical approaches to assist in determining the most effective components in a CNMA have been proposed, but the methods fall outside of the hierarchy creation framework established for standard NMA and can only be applied in connected networks \citep{seitidis_graphical_2023}. There is a need for methods of hierarchy creation that echo the framework established in standard NMA, take into account the set of components and treatments of interest, and address issues of model identifiability and parameter estimation.

In this article, we propose methodology for answering treatment and/or component hierarchy questions in CNMA.
The article is organized as follows: In Section \ref{sec:background}, we describe the general structure of popular CNMA models. In Section \ref{sec:steps}, we present methods for creating and interpreting a hierarchy in CNMA, organized into a step-by-step workflow. Of particular importance, we explain how to determine whether the relative effects of interest can be uniquely estimated from a given CNMA model and data. We illustrate the use of the proposed methods in two distinct networks in Section \ref{sec:examples}. We conclude with a discussion and recommendations for practice in Section \ref{sec:disc}.

\section{The Structure of CNMA Models} \label{sec:background}

Suppose we have the network described in Table \ref{tab:def-ex}, and we wish to analyse it using a CNMA model. Notice that the ``unit" $A$ appears alone and as a component in a multi-component treatment, units $B$ and $C$ only appear as components in multi-component treatments, unit $D$ appears only alone, and units $E$ and $F$ appear only together. A component is a unit we observe in combination with other components in two or more configurations in the network, or a unit that is only seen in combination with one or more other components. A treatment as the unit or combination of units that is employed in one or more study arms in a network. The components in this table are  $A$, $B$, and $C$. The treatments are $A$, $D$, $A+B$, $B+C$, and $E+F$. Note that although the treatment $E+F$ involves multiple units, we do not observe $E$ or $F$ in any configuration other than the treatment $E+F$, so we do not treat $E$ or $F$ as components in this network, since it would not be possible to disentangle their effects.


\begin{table}[h]
    \centering    
    \caption{Example data to illustrate the meaning of treatment and component in CNMA. Capital letters denote individual components.}
    \label{tab:def-ex}
    \begin{tabular}{c|c|c}
        Study & Treatment Arm 1 & Treatment Arm 2\\ \hline
        1 & A & B+C \\
        2 & D & B+C \\
        3 & D & A+B \\
        4 & A & E+F
    \end{tabular}
\end{table}

A CNMA model is parameterized by a vector of mean component effects, $\bbeta$, and the component effects are used to construct relative effects \citep{rucker_network_2019, wigle_bayesian_2022}. In an unanchored CNMA model, one parameter for every component, treatment not composed of components, and interaction term is estimated. In an anchored CNMA model, such as the CNMA model of \citet{rucker_network_2019} where an inactive component is specified, one of the elements of $\bbeta$ is fixed at zero, reducing the number of parameters to be estimated \citep{wigle_bayesian_2022}. We further illustrate the differences between anchored and unanchored CNMA models in the following paragraphs. More information about anchored and unanchored models is available in Supplementary Information (SI) S2.

Two matrices, the basic matrix $\BB$ and the component matrix $\CC$ are useful for understanding the available data and the assumptions in common CNMA models \citep{rucker_network_2019,li_graphical_2023}. The matrix $\BB$ describes the treatment comparisons observed in each study, and the matrix $\CC$ describes the components that make up each treatment and the structure of $\bbeta$ \citep{rucker_network_2019,li_graphical_2023}. Details of the specification of $\BB$ and $\CC$ can be found in \citep{rucker_network_2019} and \citet{li_graphical_2023}. Briefly, the $\BB$ matrix has one row for every observed comparison in the data and one column for each treatment. In an additive model, the $\CC$ matrix has one row for each observed treatment and one column for every component and remaining  treatment not made up of components. In a model with interactions, the $\CC$ matrix has columns added for each interaction term \citep{li_graphical_2023}. In an unanchored model, each row of the $\CC$ matrix has ones in the columns representing the components and interactions that make up the treatment and zeros elsewhere, to represent the fact that we estimate one element of $\bbeta$ for each component. In an anchored model, the column for the anchor or inactive component is set to all zeros, corresponding to the fact that a parameter for the anchor/inactive component need not be estimated \citep{wigle_bayesian_2022}.

To illustrate, a representation of the data shown in Table \ref{tab:def-ex} in an unanchored additive CNMA is given by
\begin{equation} \label{eq:bc}
    \BB = \begin{pNiceMatrix}[first-row,first-col]
        & A & D & A+B & B+C & E+F\\
        1 & 1 & 0 & 0 & -1 & 0 \\
        2 & 0 & 1 & 0 & -1 & 0 \\
        3 & 0 & 1 & -1 & 0 & 0 \\
        4 & 1 & 0 & 0 & 0 & -1
    \end{pNiceMatrix} \text{ and } \CC = \begin{pNiceMatrix}[first-row, first-col]
           & A & B & C & D & E+F \\
        A  & 1 & 0 & 0 & 0 & 0 \\
        D  & 0 & 0 & 0 & 1 & 0 \\
        A+B & 1 & 1 & 0 & 0 & 0 \\
        B+C & 0 & 1 & 1 & 0 & 0 \\
        E+F & 0 & 0 & 0 & 0 & 1 
    \end{pNiceMatrix}.
\end{equation}
On the other hand, an anchored additive CNMA model with treatment $D$ as the anchor has component structure described by
\begin{equation*} 
    \CC = \begin{pNiceMatrix}[first-row,first-col]
           & A & B & C & D & E+F \\
        A  & 1 & 0 & 0 & 0 & 0 \\
        D  & 0 & 0 & 0 & 0 & 0 \\
        A+B & 1 & 1 & 0 & 0 & 0 \\
        B+C & 0 & 1 & 1 & 0 & 0 \\
        E+F & 0 & 0 & 0 & 0 & 1 
    \end{pNiceMatrix},
\end{equation*}
where the entry corresponding to treatment $D$ is set to zero, corresponding to the fact that the element of $\bbeta$ related to $D$ is fixed at zero in this anchored model.
Finally, if an interaction between A and B was added to the unanchored model, the $\CC$ matrix would become
\begin{equation*}
\CC = \begin{pNiceMatrix}[first-row, first-col]
           & A & B & C & D & E+F & A:B \\
        A  & 1 & 0 & 0 & 0 & 0 & 0 \\
        D  & 0 & 0 & 0 & 1 & 0 & 0\\
        A+B & 1 & 1 & 0 & 0 & 0 & 1\\
        B+C & 0 & 1 & 1 & 0 & 0 & 0 \\
        E+F & 0 & 0 & 0 & 0 & 1 & 0
    \end{pNiceMatrix}.
\end{equation*}

The information in $\BB$ and $\CC$ can be summarised by the design matrix $\MM$, defined as
\begin{equation*}
    \MM = \BB \CC.
\end{equation*}
The design matrix corresponding with the additive unanchored model in equation \eqref{eq:bc} is given by
\begin{equation*}
    \MM = \begin{pNiceMatrix}[first-row, first-col]
        & A & B & C & D & E+F \\
        1 & 1 & -1 & -1& 0 & 0\\
        2 & 0 & -1 & -1 & 1 & 0 \\
        3 & -1 & -1 & 0 & 1 & 0 \\
        4 & 1 & 0 & 0 & 0 & -1
    \end{pNiceMatrix},
\end{equation*}
while the design matrix corresponding to the additive anchored model (with D as anchor) is given by
\begin{equation*}
    \MM = \begin{pNiceMatrix}[first-row, first-col]
        & A & B & C & D & E+F \\
        1 & 1 & -1 & -1& 0 & 0\\
        2 & 0 & -1 & -1 & 0 & 0 \\
        3 & -1 & -1 & 0 & 0 & 0 \\
        4 & 1 & 0 & 0 & 0 & -1
    \end{pNiceMatrix},
\end{equation*}
and the design matrix for the unanchored model with the interaction between components A and B is
\begin{equation*}
    \MM = \begin{pNiceMatrix}[first-row, first-col]
        & A & B & C & D & E+F & A:B \\
        1 & 1 & -1 & -1& 0 & 0 & 0\\
        2 & 0 & -1 & -1 & 1 & 0 & 0\\
        3 & -1 & -1 & 0 & 1 & 0 & -1\\
        4 & 1 & 0 & 0 & 0 & -1 & 0
    \end{pNiceMatrix}.
\end{equation*}


The design matrix determines what relative effects can be estimated from a CNMA network and model. This could include the relative effects of individual components and the relative effects of multi-component treatments that were not necessarily observed in the network \citep{rucker_network_2019}. Since additional relative effects might be estimable, additional hierarchy questions can be posed in CNMA. However, it is not always clear which relative effects can be estimated in CNMA. This presents challenges that are not yet addressed in methods for creating hierarchies in standard NMA.

\section{Methods} \label{sec:steps}

In this section, we outline the steps to creating and interpreting a valid hierarchy of treatments and/or components from a CNMA model. We also discuss any potential pitfalls in each step. There may be multiple hierarchy questions we wish to answer from a single CNMA. In that case, steps 1-4 can be repeated as necessary.

\subsection{Step 0: Select and Fit CNMA Model}

Before ranking components or treatments, one must select and fit a suitable CNMA model. We summarise some of the choices that are made in fitting a CNMA model, and their potential impacts on ranking, in Table \ref{tab:decsummary}.  In SI S1, we give an overview of some relevant characteristics of popular CNMA models.
These tables are not intended to be an exhaustive list of modelling choices or a guide on which approach to choose, but simply an overview of the factors that are most relevant to the ranking process and a summary of the standard models available at the time of writing.
Moving forward, we assume that a form of the CNMA model, that is, the set of interactions to be included, the type of model (anchored, unanchored), and the estimation approach (Bayesian, frequentist) have been chosen reasonably, and a CNMA model has been fit accordingly.

\begin{table}[h]
    \centering
    \caption{Summary of important modelling choices and their impact on the creation of component and/or treatment hierarchies.}
    \label{tab:decsummary}
    \begin{tabular}{ll}
        Decision & Impact on hierarchy creation \\ \hline
        Bayesian or frequentist & Types of models available; method for calculating ranking metrics \\
        Interaction terms included & Which relative effects can be estimated; form of relative effects \\
        Anchored or unanchored & Which relative effects can be estimated 
    \end{tabular}
\end{table}

\subsection{Step 1: Define Idealized Hierarchy Question}


A hierarchy question involves both the elements we wish to rank and the criteria we wish to use to determine ranks. An example of a hierarchy question is, ``Which component among those in the network has the highest probability of giving the best outcome?", where we want to compare all components in the network, and the criterion for preference is the probability of giving the best outcome. The criterion that we use to determine treatment preference corresponds to the ranking metric we calculate. For some ranking metrics, it is also possible to interpret the criterion in multiple ways. For example, a simple criterion would be the estimated mean relative effect versus a common reference, with larger values indicating better performance. In this case, it is clear that the ranking metric is simply the estimated mean relative effect. A more opaque example of a ranking metric is the Surface Under the Cumulative RAnking curve, or SUCRA, which corresponds to the criterion ``the fraction of competitors that a treatment beats", which can also be interpreted as ``the average rank of the treatment among its competitors", and is equivalent to using the expected ranks as the ranking metric \citep{salanti_graphical_2011, rucker_ranking_2015, salanti_introducing_2022}. For SUCRA, a larger value indicates a more preferred treatment, whereas a lower expected rank indicates a more preferred treatment. There may be interest in criteria that cannot be assessed with a standard CNMA model \citep{salanti_introducing_2022}; at this step, we assume that a criterion which can be estimated from the chosen CNMA model is selected. In Table \ref{tab:criteria}, we summarise several common ranking metrics and their interpretation in terms of a criterion for determining treatment preference.

\begin{table}[ht]
    \centering
    \renewcommand{\arraystretch}{1.3}
    \caption{Ranking metric, their interpretation as criteria for treatment preference, and the order for creating a hierarchy.}
    \label{tab:criteria}
    \begin{NiceTabular}{p{0.2\textwidth}|p{0.34\textwidth}|p{0.22\textwidth}}
        Ranking Metric & Criterion for treatment preference & Hierarchy order (most to least preferred) \\ \hline
        Point estimate of relative effects & Point estimate of relative effects & Largest to smallest\tabularnote{For a positively oriented outcome, such as number of days symptom-free.} \\
        Probability of the best value & Probability that a treatment gives the largest outcome among its competitors\tabularnote{For a positively oriented outcome, like number of days symptom-free.} & Largest to smallest \\
        Median rank & Median rank among its competitors & Smallest to largest \\
        Expected rank & Average rank among its competitors & Smallest to largest \\
        SUCRA & Fraction of competitors that a treatment beats & Largest to smallest \\
        P-score & Fraction of competitors that a treatment beats & Largest to smallest
    \end{NiceTabular}
\end{table}

In the literature, the focus in defining the treatment hierarchy question has been on the criterion, or equivalently, the ranking metric, with hierarchy questions such as ``which treatment has the largest fraction of competitors that it beats?" being posed \citep{salanti_introducing_2022}. In CNMA, we expand our focus to both the criterion for treatment performance and the set of elements to be ranked. Defining the set of treatments and/or components of interest is an important step in the creation of any hierarchy, because ranking metrics depend on the set of competing treatments used to calculate them. In standard NMA, the set of treatments that is ranked is the set of all observed treatments in the network, $\tset$. In CNMA, there are additional sets that we can consider ranking due to the potential to estimate additional effects. We can consider ranking the set of observed components based on their expected incremental effects, that is, the expected benefit when each component is combined with any other treatment. In an additive CNMA model, the incremental effect of a component $i$ is equal to $\beta_i$, the element of $\bbeta$ corresponding with component $i$ \citep{rucker_network_2019, wigle_bayesian_2022, veroniki_analysing_2025}. In CNMA models that include interactions, the benefit of combining a given component with a treatment depends on the treatment it is combined with, so we do not recommend creating a component-only hierarchy in interaction CNMA models, unless the interest is in the performance of the components used on their own. 

Another possibility presented by CNMA is to include treatments which were not observed in the data in a hierarchy, such as the combination A+C in the  example network, since the CNMA model may make it possible to estimate its effects. Ranking treatments that were not observed in the data does not present any challenge from an implementation perspective (as long as their effects are estimable, as discussed in the following step), but hierarchies that include unobserved treatments should be interpreted with extreme caution as their estimated effects are considered as exploratory or experimental \citep{rucker_network_2019}.

Finally, there may be other subsets of treatments and components that are of interest based on the the clinical context of the problem. For example, in a network consisting of treatments for depression that include psychological and pharmacological components, a particular patient may wish to select among pharmacological-only treatments. We will let $\sset$ represent a set of treatments and/or components that we are interested in ranking, which may be equal to $\tset$, a subset of $\tset$, or include additional components or treatments not included in $\tset$. One example of $\sset$ for the dataset in Table \ref{tab:def-ex} would be $\sset = \{\text{A+B, B+C, E+F}\}$, the set of all observed multicomponent treatments, which is a subset of $\tset = \{\text{A, A+B, B+C, D, E+F}\}$, the set of all observed treatments. Another example could be $\sset = \{A, A+B, B+C, A+C\}$, which is not a subset of $\tset$ because it includes a treatment not observed in the dataset. Idealized hierarchy questions could be of the form, ``what is the hierarchy of elements of $\sset$ with respect to the chosen criterion?", or ``what element of $\sset$ is the best with respect to the chosen criterion?". The hierarchy question formed in this step is idealized because it may not be possible to answer this question with the available data and CNMA model.

\subsection{Step 2: Determine Which Parameters are Estimable} \label{sec:id}

\changed{Once the idealized hierarchy question, and thus the set of interest $\sset$, is defined, it is important to ensure that all the relevant effects can be uniquely estimated by the CNMA model since ranking metrics are calculated from relative effect estimates. When the model is not identifiable, multiple values for some parameters will correspond to the same model, for example, in a standard NMA model, the value of $\bbeta$ can only be estimated up to a constant, that is, a model with $\bbeta = \bm{b}$ and a model with $\bbeta = \bm{b} + \bm{c}$ where $\bm{c}$ is a vector of the constant $c$ will correspond to the same set of relative effects between observed treatments \citep{rucker_network_2012, rucker_reduce_2014}. In other words, in standard NMA, $\bbeta$ is not uniquely estimable without arm-level modelling, but contrasts between elements of $\bbeta$, that is, relative effects between observed treatments, are uniquely estimable. In CNMA, individual elements of $\bbeta$ may become uniquely estimable due to the component structure of the model, 
but the specific elements that can be estimated will depend on the form of the CNMA model and the network of evidence. Additionally, when CNMA is used in networks that are disconnected at the treatment level, it is not guaranteed that the relative effects of all observed treatments are estimable. Since both Bayesian and frequentist CNMA models can produce estimates for parameters that cannot be uniquely estimated (the frequentist approach through the use of the pseudoinverse in least squares estimation, the Bayesian approach through the prior distribution), it is critical to assess the estimability of relative effects between the elements of $\sset$ to ensure they are meaningful before proceeding. }

\citet{li_graphical_2023} provide a graphical approach that can be used to determine which elements of $\bbeta$ can be uniquely estimated from a given network. The R package netmeta performs CNMA and will not produce relative effect estimates for contrasts that cannot be estimated \citep{balduzzi_netmeta_2023, li_graphical_2023}. Here, we build on these approaches to describe a general method for checking if an arbitrary contrast of treatments or components is uniquely estimable from a given CNMA model and network using the design matrix $\MM$. 

The set of all uniquely estimable relative effects from a model and network combination is given by the row space of the design matrix $\MM$, where the row space is defined as the set of all possible linear combinations of the rows of $\MM$. \changed{This is proven in SI S2}. For example, recall the design matrix for the additive unanchored CNMA model using the example data,
\begin{equation*}
    \MM = \begin{pNiceMatrix}[first-row, first-col]
        & A & B & C & D & E+F \\
        1 & 1 & -1 & -1& 0 & 0\\
        2 & 0 & -1 & -1 & 1 & 0 \\
        3 & -1 & -1 & 0 & 1 & 0 \\
        4 & 1 & 0 & 0 & 0 & -1
    \end{pNiceMatrix},
\end{equation*}
and suppose we are interested in the relative effect of A versus D.
The relative effect of A versus D can be represented by the row vector
\begin{equation*}
    \V{A,D} = \begin{pNiceMatrix}[first-row]
        A & B & C & D & E+F \\
        1 & 0 & 0 & -1 & 0
    \end{pNiceMatrix},
\end{equation*}
which can be obtained by taking row 1 of $\MM$ minus row 2 of $\MM$. In other words, $\V{A,D}$ is in the row space of $\MM$, and thus the relative effect of A versus D is estimable. On the other hand, the relative effect of A+C versus D is represented by
\begin{equation*}
    \V{A+C, D} = \begin{pNiceMatrix}[first-row]
        A & B & C & D & E+F \\
        1 & 0 & 1 & -1 & 0
    \end{pNiceMatrix},
\end{equation*}
which cannot be obtained from any linear combination of the rows of $\MM$, meaning that this relative effect cannot be uniquely estimated. 

It is cumbersome to determine if a vector lies in the row space of $\MM$ manually. In practice, one way to check if a row vector $\V{}$ lies in the row space of $\MM$ is to compare the rank of $\MM$ to the matrix rank of $\MM$ augmented with $\V{}$ as an additional row. If the matrix rank of $\MM$ is the same as the matrix rank of the augmented matrix, then the relative effect represented by $\V{}$ is in the row space of $\MM$ and it can be uniquely estimated, otherwise it cannot be uniquely estimated. For example, the matrix rank of $\MM$ in this example is 4. The matrix rank of the design matrix augmented with the relative effect of A+C versus D,
\begin{equation*}
    \begin{pNiceMatrix}[first-row, first-col]
        & A & B & C & D & E+F \\
        1 & 1 & -1 & -1& 0 & 0\\
        2 & 0 & -1 & -1 & 1 & 0 \\
        3 & -1 & -1 & 0 & 1 & 0 \\
        4 & 1 & 0 & 0 & 0 & -1 \\
        \V{A+C,D} & 1 & 0 & 1 & -1 & 0
    \end{pNiceMatrix},
\end{equation*}
is equal to 5. Thus, A+C versus D is not estimable in this network with the unanchored additive CNMA model. We implement this method of assessing estimability of relative effects in an R function that takes in a design matrix and a vector representing a contrast and outputs whether the contrast can be estimated, available in SI S3 and at \href{https://github.com/augustinewigle/cnmaRank}{github.com/augustinewigle/cnmaRank}.

If $\MM$ has a matrix rank equal to the length of $\bbeta$, then all elements of $\bbeta$ can be uniquely estimated, and thus relative effects of components and any multi-component treatments, observed or not, are estimable, which was shown by \citet{li_graphical_2023} and is proven in SI S2. However, this condition is not required for an arbitrary set of relative effects to be estimable. Due to the consistency assumption, it is sufficient to choose a reference treatment from $\sset$ and check that the relative effects of the remaining $\sset-1$ treatments/components versus the reference are estimable before computing ranking metrics and creating a hierarchy. 

If any relative effect is not estimable, the estimability of each element of $\bbeta$ from problematic contrasts can be assessed. For example, we can assess whether the component effect of $C$ is estimable by augmenting $\MM$ with $\V{C} = (0,0,1,0,0)$. Combinations of inestimable components may or may not result in estimable relative effects, and the only way to be sure is to check if the relative effect lies in the row space of $\MM$. Thus, we recommend only investigating which individual component effects are estimable if some or all relative effects are found to be inestimable using the method above and a better understanding of the responsible components or contrasts is needed.

\subsection{Step 3: Refine Hierarchy Question}

 In steps 1 and 2, we elucidated the set of treatments or components that we are ideally interested in, and whether or not the necessary parameters can be uniquely estimated using the CNMA model and data. If all required parameters are estimable, then the idealized hierarchy question can be answered. If not, a choice must be made. Components or treatments causing inestimable relative effects can be removed from $\sset$ to enable ranking, but this should be done with caution as it may undermine the clinical motivation for the idealized hierarchy question formulated in Step 1. For example, omitting a treatment from the ranking set may give readers the impression that the treatment is not recommended, when in reality there is simply insufficient evidence in the network, which has no relation to the actual performance of the treatment in comparison to the others. In some situations, it may be better to refrain from producing a hierarchy at all than to create a hierarchy from a reduced set of treatments or components. Practitioners must thus decide whether to revise the hierarchy question of interest by defining a reduced set of treatments or components, or to terminate the process at this step. We recommend that regardless of the choice made, practitioners be transparent about the reasons for either creating a  hierarchy from a particular subset of treatments, or for refraining from creating a hierarchy. If it is deemed appropriate to proceed with hierarchy creation, we let $\ssetf \subseteq \sset$ represent the subset of the original set of interest for which all relative effects are uniquely estimable and for which the hierarchy will be created.

\subsection{Step 4: Answer Hierarchy Question} \label{sec:calcrms}

We have now posed an answerable hierarchy question of interest, including the set $\ssetf$ and the criterion for treatment performance, which determines the ranking metric we use. We can thus proceed with the calculation of ranking metrics to create the hierarchy. In general, the procedure for calculating ranking metrics follows the same principles as in standard NMA, but the set of competing treatments is restricted to $\ssetf$ and the relative effects are calculated from the CNMA model. 

The simplest ranking metric is the point estimate of the relative effect with a common reference. Let $\hbbeta$ be an estimate of the component and interaction effects for a given CNMA model and network, and recall that in an anchored model, an element of $\hbbeta$ is fixed equal to zero.
An estimate of the relative effect of $i$ versus $j$, where $i$ and $j$ can be components or treatments, is given by taking the appropriate contrast of $\hbbeta$ estimates, and we denote it by $\hdelta{i,j}$. In particular, if the desired relative effect of $i$ versus $j$ is coded in the row vector $\V{i,j}$ as described in Section \ref{sec:id}, the relative effect estimate is given by
\begin{equation} \label{eq:calcRE}
    \hdelta{i,j} = \V{i,j}\hbbeta.
\end{equation}
To create a hierarchy that can answer questions relating to the criterion ``estimated relative effect size", choose a common reference treatment or component $r$ in $\ssetf$, calculate $\hdelta{i,r}$ for all $i$ in $\ssetf$, and order the elements of $\ssetf$ by the relative effects. 

The previous ranking metric does not consider any uncertainty in the estimation of the relative effects. Other criteria, like ``probability of having the best value" or ``fraction of competitors a treatment is expected to beat" incorporate some consideration of uncertainty in the relative effects. For the following methods, we assume that we have available $M$ samples of $\hbbeta$, such as through MCMC sampling in fitting a Bayesian CNMA model, or by resampling from the sampling distribution of $\hbbeta$ in a frequentist CNMA model \citep{wigle_bayesian_2022, rucker_ranking_2015}. For each sample, relative effects to a common reference $r$ are calculated as in Equation \ref{eq:calcRE} for all elements of $\ssetf$, given by $\hdeltam{i,r}$ for all $i$ in $\ssetf$ and $m = 1$, $\dots$, $M$. In the remainder of this section, we describe how to use these samples to calculate the remaining ranking metrics in Table \ref{tab:criteria} in a setting where the outcome is oriented so that a larger response is preferred, such as symptom-free days. The extension to a negatively-oriented outcome, such as mortality, is straightforward.

The probability that treatment or component $i$ gives the best value out of the competitors in $\ssetf$ is given by
\begin{equation}
    \Pbest(i) = \frac{1}{M} \summ I\left(\hdeltam{i,r} > \hdeltam{j,r} \text{ for all }j\neq i \in \ssetf\right),
\end{equation}
where $I(\cdot) = 1$ when its argument is true and 0 otherwise.

A simple method for determining the median and expected ranks is to first determine the rank in each sample. The rank of treatment $i$ out of set $\ssetf$ in the $m^{\text{th}}$ sample is given by
\begin{equation}
     \rankm(i) = |\ssetf| - \sumneq I\left(\hdeltam{i,r} > \hdeltam{j,r}\right),
\end{equation}
where $|\ssetf|$ is the number of elements in $\ssetf$.
The median rank of treatment $i$ can then be found by ordering $\rankm(i), m = 1,$ $\dots$, $M$ from smallest to largest and selecting the middle value, or the average of the middle values when $M$ is even. The expected rank can be found by averaging the ranks, that is,
\begin{equation}
    \E\left[\rank(i)\right] = \frac{1}{M}\summ \rankm(i).
\end{equation}
The SUCRA value for treatment or component $i$ in set $\ssetf$ can easily be calculated from the expected rank via
\begin{equation}
    \sucra(i) = \frac{|\ssetf| - \E\left[\rank(i)\right]}{|\ssetf|-1},
\end{equation}
making clear the equivalence between these ranking metrics/criteria \citep{rucker_ranking_2015}. 

P-scores are a frequentist equivalent to SUCRAs which do not use the $M$ samples, instead relying on the assumption that the sampling distribution of $\hbbeta$ is normal, thus only requiring the relative effect point estimates and their standard errors, which can be obtained for arbitrary relative effects in R package netmeta using the function netcomparison \citep{rucker_ranking_2015,balduzzi_netmeta_2023}. The P-score of treatment or component $i$ in set $\ssetf$ is given by
\begin{equation}
    \Pbar(i) = \frac{1}{|\ssetf|-1}\sumneq \Phi\left(\frac{\hdelta{i,j}}{\sigma_{ij}}\right),
\end{equation}
where $\sigma_{ij}$ is the standard error of $\hdelta{i,j}$ and $\Phi$ represents the cumulative distribution function of the standard normal distribution \citep{rucker_ranking_2015}.

The hierarchy question is then answered by ordering the treatments and/or components by their ranking metrics as described in Table \ref{tab:criteria}. Each treatment or component hierarchy must be interpreted in the context of the set of compared treatments/components $\ssetf$, and ranking metrics (besides the point estimates) should not be compared between different sets of ranked treatments since their values depend on the size and membership of $\ssetf$. To avoid exaggerating small differences between treatments or components, we agree with the recommendation of reporting the ranking metric values, point estimate and confidence or credible intervals for the relative effects alongside the hierarchy rankings \citep{hutton_prisma_2016}.

\section{Case Studies} \label{sec:examples}

We now illustrate the methods in two distinct case studies based on published networks. In each case study, we consider two hierarchy questions, thus we follow the steps twice for each network. We used the R package netmeta to fit frequentist CNMA models in both cases \citep{balduzzi_netmeta_2023}. To calculate ranking metrics that require resampling, we took 1000 independent samples from $N(\hdelta{i,r}, \sigma_{ir})$ for each treatment $i$ in $\ssetf$, where $r$ is the first treatment or component in $\ssetf$. Code to reproduce both case studies is available at \href{https://github.com/augustinewigle/cnmaRank/}{github.com/augustinewigle/cnmaRank}.

\subsection{Case Study 1: Primary Care of Depression}

\begin{figure}
    \centering
    \includegraphics[width=0.7\linewidth]{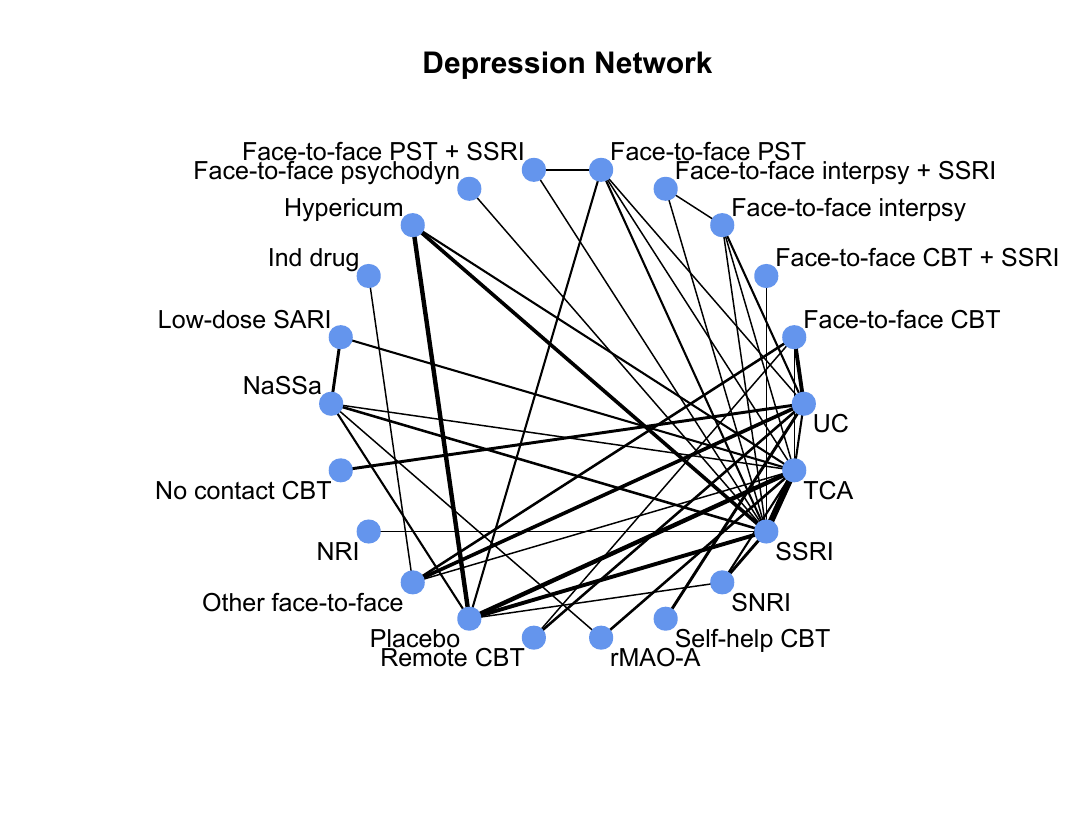}
    \caption{Network of treatments for primary care of depression \citep{linde_questionable_2016}.}
    \label{fig:netdepress}
\end{figure}

In this case study, we use a dataset consisting of 93 studies of treatments for depression in primary care, available in R package netmeta and originally described by \citet{linde_questionable_2016}. The network compares 22 treatments, including placebo, pharmacological, psychological, and combination interventions. In particular, three multi-component treatments, face-to-face Cognitive Behavioural Therapy (CBT) + Selective Serotonin Reuptake Inhibitors (SSRI), face-to-face Problem Solving Therapy (PST) + SSRI, and face-to-face interpersonal psychotherapy (interpsy) + SSRI, are present in the network, with each of their components also observed as single component treatments. The network is connected at the treatment level, and is visualized in Figure \ref{fig:netdepress}. 

\subsubsection{Step 0}
A random-effects additive CNMA model was fit, with no inactive components specified, corresponding to an unanchored model \citep{wigle_bayesian_2022, rucker_network_2019}. The outcome is response to treatment, defined as at least a 50\% reduction in the score on a depression symptom severity scale from baseline, with the odds ratio used as the effect measure.

\subsubsection{First Hierarchy Question}\label{sec:firstdepressq}

\subsubsubsection{Step 1: Idealized Hierarchy Question}

For our first hierarchy question, we would like to provide a hierarchy of all treatments observed in the network to help summarise the CNMA results. We thus define $\sset = \tset$, the set of all 22 observed treatments in the network. For the preference criterion, we select the median rank among $\sset$. The idealized hierarchy question is, ``what is the hierarchy of all 22 observed primary care treatments for depression in terms of their median ranks?".

\subsubsubsection{Step 2: Determine Which Parameters are Estimable}

To assess whether relative effects are estimable, we choose face-to-face CBT as the reference $r$, and use the augmented rank comparison method described in Section \ref{sec:id} to check the estimability of the relative effect of all 22 treatments in $\sset$ versus face-to-face CBT. The design matrix $\MM$ for this network is of dimension $124\times 19$, and can be seen in the R code on Github. The rank of $\MM$ is 19, which is equal to the length of $\bbeta$. We thus find that all relative effects are estimable.

\subsubsubsection{Step 3: Refine Hierarchy Question}

Since all relative effects for elements of $\sset$ are estimable, we do not need to refine the idealized hierarchy question. Let $\ssetf = \sset$.

\subsubsubsection{Step 4: Answer Hierarchy Question}

\begin{figure}[ht]
    \centering
    \includegraphics[width=0.8\linewidth]{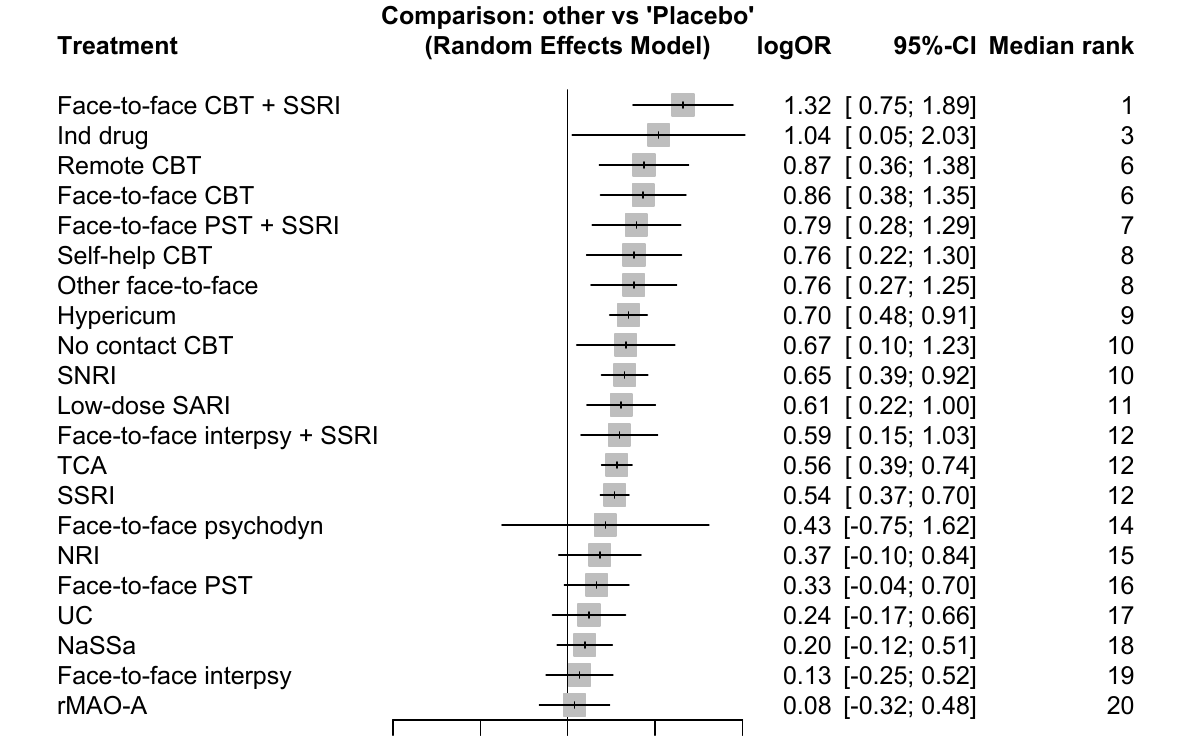}
    \caption{Forest plot showing estimated relative effects of all treatments in the depression network versus placebo and median ranks to answer the hierarchy question in Section \ref{sec:firstdepressq}.}
    \label{fig:depress}
\end{figure}

We answer the hierarchy question by computing median ranks for $\ssetf$ as described in Section \ref{sec:calcrms}. A forest plot of the estimated log odds ratio (OR) of each treatment in $\ssetf$ compared to placebo, the 95\% confidence interval (CI), and the median rank is shown in Figure \ref{fig:depress}, with treatments ordered from most to least preferred. The hierarchy from the median ranks agrees with the ordering of the point estimates of the relative effects, although there are ties in the median ranks whereas ties in the point estimates are only due to rounding.

\subsubsection{Second Hierarchy Question} \label{sec:seconddepressq}

\subsubsubsection{Step 1: Idealized Hierarchy Question}

The second hierarchy question concerns the incremental effects of the components. The components are face-to-face CBT, face-to-face interpsy, face-to-face PST, and SSRI, which we define as $\sset$. In this context, the incremental effect of a component can be interpreted as the change in the log OR of response when the component is added to another treatment, which we assess via the probability of giving the largest change (best value). Our hierarchy question is, ``Which component has the highest probability of giving the highest increase in the log odds of response when added to an arbitrary treatment?".

\subsubsubsection{Step 2: Determine Which Parameters are Estimable}

As all four components are observed as single component treatments in the network, and we already determined that all elements of $\bbeta$ are uniquely estimable in Section \ref{sec:firstdepressq}, it is clear that all relative effects in $\sset$ are also estimable.

\subsubsubsection{Step 3: Refine Hierarchy Question}

Once again, since all relative effects for elements of $\sset$ are estimable, we can proceed with our idealized hierarchy question, and let $\ssetf = \sset$.

\subsubsubsection{Step 4: Answer Hierarchy Question}

We calculated the probability of having the best value for each component in $\ssetf$ as described in Section \ref{sec:calcrms}. The results are summarised in Table \ref{tab:depress}. We found that face-to-face CBT has the highest probability of having the best incremental effect, with a value of 0.826. This represents a much higher probability of giving the best value compared to the other components, with the next highest probability being 0.162 for SSRI.

\begin{table}[ht]
    \centering
    \caption{Point estimate and 95\% confidence interval (CI) for relative effects of components of interest in section \ref{sec:seconddepressq}, probability of having the best value, and resulting hierarchy.}
    \label{tab:depress}
    \begin{tabular}{l|l|l|l}
        Component & $\hdelta{i,\text{Face-to-face CBT}}$ (95\% CI) & $\Pbest(i)$&  Hierarchy \\ \hline
        Face-to-face CBT & 0 (-,-) & 0.826 &1 \\
        SSRI & -0.327 (-0.800, 0.146) & 0.162 & 2\\
        Face-to-face PST & -0.533 (-1.06, -0.005) & 0.010 & 3\\
        Face-to-face interpsy & -0.730 (-1.224, -0.236) & 0.002 & 4
    \end{tabular}
\end{table}

\subsection{Case Study 2: Treatment of Relapse or Refractory Chronic Lymphocytic Leukemia} \label{sec:casestudy2}

\begin{figure}
    \centering
    \includegraphics[width=0.7\linewidth]{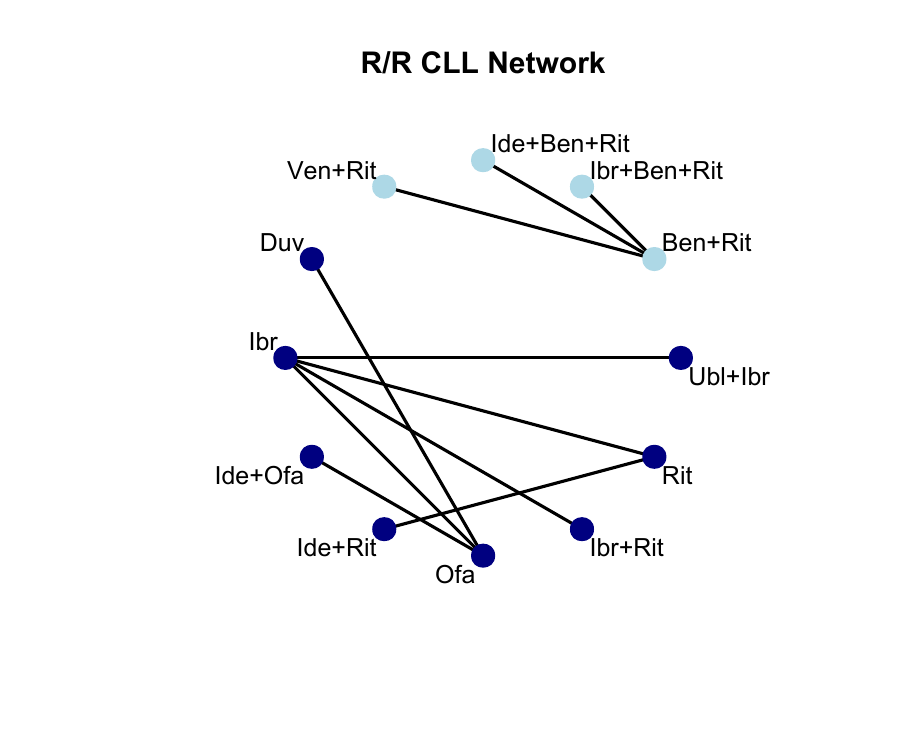}
    \caption{Network of targeted agents for the treatment of refractory/relapse chronic lymphocytic leukemia (R/R CLL) \citep{li_graphical_2023, chen_treatment_2019}.}
    \label{fig:netcll}
\end{figure}

We illustrate ranking in CNMA using a set of RCTs concerning novel targeted agents for relapse or refractory chronic lymphocytic leukemia (R/R CLL) \citep{li_graphical_2023, chen_treatment_2019}. The data are freely available from the Supplement to \citet{li_graphical_2023}. The data consist of 10 RCTs that cover 12 unique treatments made up of eight components: Bendamustine (Ben), Duvelisib (Duv), Ibrutinib (Ibr), Idelalisib (Ide), Ofatumumab (Ofa), Rituximab (Rit), Ublituximab (Ubl), and Venetoclax (Ven). The RCTs form a network that is disconnected at the treatment level, forming the light blue and dark blue subnetworks as shown in Figure \ref{fig:netcll}, but can be analysed using a CNMA model due to common components between the subnetworks. The components Ben, Ofa, and Rit are considered traditional therapy for R/R CLL, while the remaining components represent novel targeted agents \citep{chen_treatment_2019}. The network is quite sparse, as it is disconnected at the treatment level and there are no two studies that made the same treatment comparison.

\subsubsection{Step 0}

The reported outcome is the log of the progression-free survival hazard ratio, so larger values indicate a better treatment \citep{li_graphical_2023, chen_treatment_2019}. We chose to fit a common-effect unanchored CNMA model without interaction terms.

\subsubsection{First Hierarchy Question} \label{sec:firstcllq}

\subsubsubsection{Step 1: Idealized Hierarchy Question}

Similarly to the depression case study, our first hierarchy question concerns all observed treatments in the network. We define $\sset = \tset$, the set of all 12 treatments in the network. We will use the point estimate of the relative effects as the criterion for preference. The idealized hierarchy question is, ``What is the hierarchy of the observed treatments in terms of their estimated relative effects?".

\subsubsubsection{Step 2: Determine Which Parameters are Estimable }

We assess whether the relative effects can be estimated using the augmented matrix rank comparison method for the set $\sset$, with Ben + Rit as the reference.
The design matrix of this network is given by
\begin{equation*}
    \MM=
    \begin{pNiceMatrix}[first-row]
         Ben & Duv & Ibr & Ide & Ofa & Rit & Ubl & Ven \\ 
0 & 0 & 1 & 0 & -1 & 0 & 0 & 0 \\ 
  0 & 0 & 0 & 0 & 0 & 0 & -1 & 0 \\ 
  0 & 0 & 0 & 1 & 0 & 0 & 0 & 0 \\ 
  0 & 1 & 0 & 0 & -1 & 0 & 0 & 0 \\ 
  0 & 0 & -1 & 0 & 0 & 0 & 0 & 0 \\ 
  1 & 0 & 0 & 0 & 0 & 0 & 0 & -1 \\ 
  0 & 0 & 0 & -1 & 0 & 0 & 0 & 0 \\ 
  0 & 0 & 1 & 0 & 0 & -1 & 0 & 0 \\ 
  0 & 0 & 0 & 1 & 0 & 0 & 0 & 0 \\ 
  0 & 0 & 0 & 0 & 0 & -1 & 0 & 0 \\ 
    \end{pNiceMatrix}.
\end{equation*}
We find that none of the relative effects with Ben + Rit as reference are estimable. To understand the source, we check the estimability of all component effects individually, and find that Ben and Ven are not estimable while all other components are. 

\subsubsubsection{Step 3: Refine Hierarchy Question}

In the previous step, we found that components Ben and Ven are not estimable. Revising our set of interest to exclude any treatments that involve these components would require removing four out of 12 total treatments, and amounts to ignoring the light blue subnetwork in Figure \ref{fig:netcll} in ranking. Since this could give the impression that these treatments are not effective compared to the others, and reduces the set of interest by a third of the treatments, we refrain from refining the hierarchy question and end the process here.

\subsubsection{Second Hierarchy Question}\label{sec:secondcllq}

\subsubsubsection{Step 1: Idealized Hierarchy Question}

In an initial analysis of the larger subnetwork, it was found that novel targeted therapies were superior to the traditional treatment Ofa \citep{chen_treatment_2019}. This motivates us to consider the performance of the novel targeted therapy components, that is, define $\sset = \{\text{Duv, Ibr, Ide, Ubl, Ven}\}$. We will rank the components based on their expected rank. The idealized hierarchy question is ``Which of the novel targeted therapy components has the highest expected rank?".

\subsubsubsection{Step 2: Determine Which Parameters are Estimable }

Assessing estimability of each of the components relative to the common reference Duv, we find that the relative effect of Ven versus Duv is inestimable, as expected based on our findings in Section \ref{sec:firstcllq} Step 2. The relative effects of other components in $\sset$ are estimable.

\subsubsubsection{Step 3: Refine Hierarchy Question}

We found in the previous step that Ven cannot be ranked alongside the other elements of $\sset$. One way to proceed is to create a hierarchy of a reduced set of components, $\ssetf = \{\text{Duv, Ibr, Ide, Ubl}\}$. The refined hierarchy question corresponding to $\ssetf$ is ``Which of the novel targeted therapy components (besides Ven) has the highest expected rank?". In reporting the results, we will emphasize that Ven was excluded from the hierarchy for technical reasons, not because it is not effective, and call for additional studies that may make relative effects of Ven estimable to enable ranking.

\subsubsubsection{Step 4: Answer Hierarchy Question}

We calculated expected ranks for each component in $\ssetf$ as described in Section \ref{sec:calcrms}. The results are shown in Table \ref{tab:cll}. Since we initially wanted to include Ven in the hierarchy, we included a row in the table for this component to emphasize that it is relevant from a clinical perspective. We find that Duv has the highest expected rank out of Duv, Ibr, Ide, and Ubl. Ven is excluded from the hierarchy because its relative effects cannot be estimated from the present model and data.

\begin{table}[ht]
    \centering
    \caption{Point estimate and 95\% confidence interval (CI) for relative effects of components of interest in Section \ref{sec:secondcllq}, expected rank, and resulting hierarchy. Note that Ven was included in the idealized hierarchy question, but could not be ranked as its relative effects are not estimable.}
    \label{tab:cll}
    \begin{tabular}{l|l|l|l}
        Component & $\hdelta{i,Duv}$ (95\% CI) & $\E\left[\rank(i)\right]$ out of 4 & Hierarchy \\ \hline
        Duv & 0 (-,-)& 1.158 & 1\\
        Ubl & -0.548 (-1.888, 0.792) & 2.033 & 2\\
        Ide & -1.314 (-2.230, -0.397)& 3.051 & 3\\
        Ibr & -1.609 (-2.336, -0.883)& 3.758 & 4\\
        Ven & Not estimable & - & -\\
    \end{tabular}
\end{table}

\section{Discussion and Conclusions} \label{sec:disc}

Treatment hierarchies are popular outputs of NMAs. CNMA is a useful extension of NMA for networks containing multi-component treatments. In this article, we provided a structured workflow for asking and answering hierarchy questions about treatments and components using standard CNMA models. Of particular importance, we clarified what relative effects can be estimated from a given CNMA network and model, and how to check if a set of relative effects are estimable in practice. We showed the versatility and applicability of the proposed workflow in two case studies.

The methods described in this article focused on CNMA models that share a similar structure. More complex CNMA models have been described, including an approach that uses individual participant data \citep{efthimiou_bayesian_2022} and one designed for networks of interventions that can be coded as a set of shared features \citep{davies_complex_2024}. Extending the methods presented here to more intricate CNMA models is a potential area of future work. 

One reason for using a CNMA model is to reconnect a disconnected network, as in the case study in Section \ref{sec:casestudy2}. We demonstrated that reconnecting the network through common components does not guarantee meaningful effect estimates or treatment rankings. Although problems with estimation are more likely in disconnected networks, they can arise in connected networks as well. This underscores the importance of Step 2, assessing estimability, in the proposed workflow. In addition, the graphical approach described by \citet{li_graphical_2023} could be used in conjunction with our proposed methods to determine what additional data would be needed to enable hierarchy creation for a desired set $\sset$. 

\changed{In this work, we used a definition of identifiability and corresponding methods that relate to the likelihood of the data. However, other notions of identifiability and estimability that take into account prior distributions can be applied in Bayesian models, such as the degree of overlap between prior and posterior distributions, or the ratio of their variances \citep{garrett_latent_2000,xie_measures_2006}. In the context of NMA, \citet{beliveau_theoretical_2021} derived asymptotic results for the ratio of the posterior and prior variances of relative effects in disconnected networks, and showed that using informative priors can improve Bayesian learning in some situations. Outside of the latter work, little attention has been paid to Bayesian learning and identifiability in NMA. Further investigation into identifiability in Bayesian NMA and CNMA models, and the impact of using informative priors, may uncover new insights that enable meaningful estimation of more relative effects.}

Finally, we emphasize that treatment hierarchies are meant to supplement the resulting point estimates and CIs of relative effect estimates, not replace them. We have shown examples of how treatment and component hierarchies can be presented accurately in the case studies. Different ranks in the hierarchy do not necessarily reflect clinically or statistically significant differences between treatments or components. Extensions of SUCRA and P-scores that incorporate clinically important differences \citep{curteis_ranking_2025, mavridis_extensions_2020} could be applied in CNMA to create hierarchies that are more clinically relevant in the future. Additionally, the use of other ranking metrics or ranking approaches are possible, for example, \citet{ades_treatment_2025} provided a framework for treatment recommendations that uses a loss function and explicitly considers uncertainty. These procedures could be extended to the case of CNMA by using the principles described in Section \ref{sec:calcrms} to calculate relative effects.

\section{Acknowledgments}

AW acknowledges the support of the Natural Sciences and Engineering Research Council of Canada (NSERC), Grant numbers CGS-D 569445-2022, CGS-MSFSS 588256-2023, PDF 598932-2025. LL was supported in part by the US National Institute on Aging grant R03 AG093555, the US National Library of Medicine grant R21 LM014533, and the Arizona Biomedical Research Center grant RFGA2023-008-11. The content is solely the responsibility of the authors and does not necessarily represent the official views of the US National Institutes of Health and the Arizona Department of Health Services.

\bibliographystyle{unsrtnat}
\bibliography{awigle-freiburg}

\end{document}